\documentclass[a4paper,11pt]{article}
\pdfoutput=1

\usepackage[utf8]{inputenc}
\usepackage[margin=1in]{geometry}
\usepackage[T1]{fontenc}
\usepackage{booktabs,multirow}
\usepackage{bbold}
\usepackage{subcaption}
\usepackage{amsmath}
\usepackage{mathtools,amssymb,braket,dsfont}
\usepackage{url}
\usepackage{graphicx}
\usepackage{multicol}
\usepackage{multirow}
\usepackage{float}
\usepackage[dvipsnames]{xcolor}
\usepackage{slashed}
\usepackage{jheppub} 

\hypersetup{
	colorlinks=true,
	linkcolor=blue,
	citecolor=blue,
	filecolor=magenta,
	urlcolor=blue
}
\usepackage[numbers, sort&compress]{natbib}
\usepackage[utf8]{inputenc}
\usepackage[T1]{fontenc}
\usepackage{comment}
\usepackage{orcidlink}

\usepackage{enumitem}

\preprint{P3H-25-058, TTP25-027}

\title{TeV-scale scalar leptoquarks motivated by B anomalies improve Yukawa unification in SO(10) GUT}

\author[a]{Xiyuan Gao\orcidlink{0000-0002-1361-4736},}

\author[a]{Ulrich Nierste\orcidlink{0000-0002-7442-4776}}

\affiliation[a]{Institute for Theoretical Particle Physics, Karlsruhe Institute of Technology (KIT),\\Wolfgang-Gaede-Str. 1, D-76131 Karlsruhe, Germany}

\emailAdd{xiyuan.gao@kit.edu}
\emailAdd{ulrich.nierste@kit.edu}

\abstract{
It is common practice to explain deviations between data and Standard-Model (SM) predictions by postulating new particles at the TeV scale ad-hoc. This approach becomes much more convincing, if one successfully embeds the postulated particles into a UV completion which addresses other conceptual or phenomenological shortcomings of the SM. We present a study of an SO(10) grand unified theory which contains scalar leptoquark fields employed to explain the ``flavour anomalies'' in $b\rightarrow s$ and $b\rightarrow c$ decays. We find that the additional degrees of freedom improve the renormalization-group (RG) evolution of the SM parameters. In particular, the light leptoquarks modify the RG evolution of the Yukawa couplings such that successful bottom-tau unification becomes possible in a minimal SO(10) GUT with only a  $126$-plet coupling to fermions. If we amend the Yukawa interaction of the minimal one-generation model with a second fermion multiplet and small flavor-violating terms, we find the flavour violation in the leptoquark couplings growing with the RG evolution while it stays small in the Yukawa interaction of the SM Higgs boson. By employing mass splittings among the members of the 126 -plet one can increase the effect and obtain large flavor violation in leptoquark couplings from tiny perturbations at the GUT scale, because the flavour-conserving limit is an unstable initial condition for the RG equations. }

\begin{document}

\maketitle
\flushbottom


\section{Introduction}

Grand unification theories (GUTs)~\cite{Georgi:1974sy, Fritzsch:1974nn} offer an appealing framework for new physics beyond the Standard Model (SM). One of their key successes is
the explanation of the quantization of hypercharge in units of the weak isospin, which implies the quantization of electric charge in a way that neutron and neutrinos are electrically neutral.
 No fundamental principles within the SM can forbid deviations from this pattern \cite{Foot:1992ui,Babu:1989tq}, but it has been tested to extremely high precision; for instance, the neutron charge is constrained to be smaller than about $10^{-21}~e$~\cite{Bressi:2011yfa}. This puzzle reflects the arbitrariness in constructing anomaly-free representations of the SM gauge group $G_{\text{SM}}=SU(3)_C\times SU(2)_L\times U(1)_Y$, while, by contrast, the quantum numbers of SM fermions do not appear random in reality, but directly point to an extended symmetry: Each generation, with a right-handed neutrino, fits neatly into the spinor representation $16_F$ of SO(10). `Unifiability' is a rare feature among all possible anomaly-free assignments of $G_{\text{SM}}$ representation~\cite{Herms:2024krp}, and strongly suggests that the fundamental gauge group of nature is a single semi-simple group $G_{\text{GUT}}$, such as SU(5) or SO(10) containing $G_{\text{SM}}$ as a subgroup. Moreover, various fits to gauge coupling unification for SO(10) suggest that the GUT breaking scale $M_{\text{GUT}}$ lies around $10^{16}$ GeV~\cite{Deshpande:1992au, Deshpande:1992em, Bertolini:2009qj, Ohlsson:2019sja}, in principle 
 making the GUT idea testable by future proton decay experiments~\cite{Hyper-Kamiokande:2018ofw, Dev:2022jbf}. 
GUTs predict that gauge coupling unify. With the measured values of the SM gauge 
couplings one finds that this feature holds qualitatively, as the couplings converge to each 
other at high energies, but fails quantitatively \cite{Amaldi:1991cn, Langacker:1991an}. Quantitative gauge coupling unification can be achieved in multiple ways by "populating the desert" between the electroweak scale and $M_{\text{GUT}}$. For example, in supersymmetric GUTs the superpartners of the SM particles with ${\cal O}(1-100~\text{TeV})$ masses make gauge unification possible, but one can also 
employ mass splittings among the members of the large Higgs multiplets to modify the renormalization group (RG) evolution of the couplings near $M_{\text{GUT}}$.

This paper addresses the unification of Yukawa couplings. This topic has two key elements,
the number of Higgs multiplets coupling to fermions and the light degrees of freedom invoked to 
change the RG evolution of the Yukawa couplings between the electroweak scale and $M_{\text{GUT}}$.
Unfortunately, the most economical SU(5) or SO(10) Yukawa sectors fail in reproducing the observed fermion masses and mixing angles.
It is instructive to study the wrong prediction of the $b-\tau$ mass relation. New particles are required to achieve realistic fermion masses and mixings --- either heavier than $M_{\text{GUT}}$ (entering as effective operators~\cite{Ellis:1979fg, Preda:2022izo, Preda:2024vas}) or lighter (such as additional scalars containing Higgs doublets~\cite{Georgi:1979df, Bajc:2002iw, Babu:2016bmy, Preda:2025afo} or vector-like fermions~\cite{Dorsner:2014wva, Babu:2016cri}). The predictions of the \textit{most minimal} GUTs are replaced by fits within \textit{minimal realistic} models which contain additional free parameters; see Refs.~\cite{Bajc:2005zf, Joshipura:2011nn, Dueck:2013gca, Babu:2016bmy, Ohlsson:2019sja, Mummidi:2021anm, Chen:2021zwn, Patel:2022xxu, Haba:2023dvo, Chen:2024cht} for examples of non-supersymmetric GUTs.
Here with \textit{most minimal} we mean GUTs in which only one Higgs multiplet couples to fermions, in the case of SO(10) the corresponding representations can be $10_H$, $120_H$, or $126_H$. 
As a common feature of the proposed minimal realistic models, many robust and discriminative predictions of the \textit{most minimal} GUT are lost.  

However, it would be premature to claim that \textit{most minimal} GUTs are unrealistic: The loophole is the assumption of  a particle desert between the electroweak and GUT scales, usually deduced from naturalness criteria~\cite{Georgi:1979md}. 
The impact of the deviation of the desert picture on Yukawa unification is best studied in
supersymmetric GUTs, in which the infrared (IR) theory is the Minimal Supersymmetric Standard Model (MSSM). The threshold correction to the matching relation between SM and MSSM Yukawa couplings can be enhanced when $\tan\beta$, the ratio of the vacuum expectation values (VEVs) of the two MSSM Higgs doublets, is large~\cite{Hall:1993gn, Carena:1993ag, Carena:1999py, Diaz-Cruz:2000nvf, 
Girrbach:2011an, Deppisch:2018flu}. 
This fact changes the predictions of minimal GUT, and together with successful gauge coupling unification, has been viewed as indirect support for low-scale supersymmetry. However, the SUSY GUT is far from minimal. It requires many more physical particles than the non-SUSY one, and the huge number of physical degrees of freedom brings a new puzzle that perturbative expansion could fail~\cite{Milagre:2024wcg, Bajc:2016efj}. Recently, the authors of Ref.~\cite{Patel:2023gwt,Shukla:2024bwf} proposed a new idea, that the wrong $b-\tau$ mass relation in minimal SU(5) can be resolved by introducing a large mass hierarchy among the particles within the same scalar multiplet. This hierarchy requires that some particles lie far below the $M_{\text{GUT}}$, and suggests the desert picture together with the naturalness criterion should be reconsidered. The wrong fermion mass pattern does not necessarily falsify the minimal GUT; rather, it implies that SM alone cannot serve as a viable IR theory. 
Some scalar particles from the GUT sector need to be included in the light spectrum. In some cases, such particles also lead to successful gauge coupling unification; see, for instance, Ref.~\cite{Dorsner:2005fq, Bajc:2006ia, Preda:2022izo, Goto:2023qch, Preda:2024vas}.

From a different  perspective, doubts on the particle desert picture 
are nurtured by experimental data on flavor-changing $B$ meson decays. For 
more than a decade several observables related to $b\to s \mu^+\mu^-$ or 
$b\to c \tau\nu$ decays have been found to deviate form their SM predictions. The current status of the "flavor anomalies" is as follows: $b\to c \tau\nu$ is probed through the ratios of branching ratios $R(D^{(*)})=\text{BR}(B\to D^{(*)} \tau \nu)/\text{BR}(B\to D^{(*)} \ell 
\nu)$ ($\ell=e,\mu)$ and polarisation data with very robust theory predictions, because 
the only non-perturbative quantity involved is a ratio of form factors multiplying a term suppressed by the mass ratio $m_\tau^2/m_B^2$.  An analysis exploiting experimental information on form factor shapes finds the combined $b\to c\tau\nu$ data deviating from the SM predictions by 4.4$\sigma$ \cite{Iguro:2024hyk}. A recent paper calculating form factors from first principles finds compatible results with slightly larger theory uncertainties. Moreover, if one tried to 
change the form factor ratio in $R(D^{*})$ to a level that the data are reproduced, 
predictions of measured polarisation data 
in $B\to D^* \ell\nu$ decays with light leptons $\ell=e,\mu$ would instead severely deviate from their SM predictions \cite{Fedele:2023ewe}.
BaBar, Belle, Belle II and LHCb contribute to the $b\to c\tau\nu$ anomaly with mutually consistent measurements \cite{BaBar:2012obs,BaBar:2013mob,Belle:2019rba,LHCb:2023cjr,Belle:2015qfa,Belle:2016dyj,Belle:2017ilt,Belle-II:2025yjp} (combined in  Ref.~\cite{HFLAV:2022esi}). 
The $b\to s \mu^+\mu^-$ anomaly is supported by measurements of various branching ratios and angular distributions of $b$-flavored hadrons by LHCb~\cite{LHCb:2015ycz,LHCb:2015wdu,ATLAS:2018cur,CMS:2018qih, LHCb:2018jna, CMS:2019bbr,BELLE:2019xld,CMS:2020oqb,LHCb:2020dof,LHCb:2020gog, LHCb:2021vsc,LHCb:2021trn,LHCb:2021xxq,LHCb:2021lvy, LHCb:2021zwz, LHCb:2022qnv,LHCb:2022zom}
and, more recently, also by CMS \cite{CMS:2024atz}. Moreover, all data are compatible with 
effects of equal size in $b\to s e^+ e^-$,i.e.\ lepton-flavor universality in the first two generations \cite{LHCb:2022qnv}. 
The combination of all data prefers beyond-SM (BSM) scenarios with a significance above 5$\sigma$ \cite{Capdevila:2023yhq} if the SM prediction of  \cite{Khodjamirian:2010vf} is used. The 
latter has been challenged by several alternative calculational approaches \cite{Bobeth:2017vxj,Gubernari:2023puw,Bordone:2024hui,Isidori:2024lng,Isidori:2025dkp,Hurth:2025vfx} and while more conservative estimates of hadronic uncertainties reduce the significance of BSM physics, there is no convincing way to bring the data into good agreement with the SM predictions. Statistically, it is unlikely that all these anomalies will disappear in the future~\cite{Crivellin:2023zui}, and their BSM explanation requires particles not far above the TeV scale. Specifically, leptoquarks (LQs) with masses between 1 TeV and 50 TeV are well-suited to remedy the flavor anomalies without harming predictions of observables which are in agreement with their SM predictions \cite{Sakaki:2013bfa,Dorsner:2016wpm,Dumont:2016xpj,Li:2016vvp,Bhattacharya:2016mcc,Chen:2017hir,Crivellin:2017zlb,Cai:2017wry,Jung:2018lfu,Aydemir:2019ynb,Popov:2019tyc,Crivellin:2019dwb,Bigaran:2019bqv,Bansal:2018nwp,Iguro:2020keo,Ciuchini:2022wbq,Dev:2024tto,Fedele:2023rxb,Bigaran:2024vnl}. 

The state-of-the art is to postulate the required light LQs ad-hoc, which remains unsatisfactory until these particles are embedded into a meaningful 
theory addressing fundamental puzzles of the SM. Thus it is a natural idea to analyze whether the (multi-)TeV scale leptoquarks are beneficial to GUTs, as we do in this paper. 
Although LQs can arise naturally in many partial unification frameworks, such as Pati-Salam (PS) theories \cite{Pati:1974yy}, their masses and interactions with fermions remain puzzling. If the LQs are vector bosons, their masses originate from  spontaneous gauge symmetry breaking and are their mass is protected by gauge symmetry. 
However, if PS unification is realized at the multi-TeV scale, the LQ-fermion coupling structure needs to respect an approximate $U(2)^n$ flavor symmetry to evade the bounds from processes involving light flavors, such as $K_L\rightarrow \mu e$~\cite{Barbieri:2011ci, Isidori:2012ts, Barbieri:2012uh,Blankenburg:2012nx, Faroughy:2020ina, Antusch:2023shi}. Additional vector-like Fermions and/or extended gauge groups are typically needed to achieve TeV-scale partial unification while preserving $U(2)^n$ at low energies~\cite{DiLuzio:2017vat, Bordone:2017bld, Greljo:2018tuh,Calibbi:2017qbu, Fuentes-Martin:2022xnb, Davighi:2022bqf}. On the other hand, \emph{scalar}\ LQs can naturally preserve the chiral $U(2)^n$ symmetry, but their masses are unprotected and suffer from the same fine-tuning problem as the SM-Higgs boson~\cite{PhysRevD.27.1601}. This conceptual puzzle becomes an explicit problem in a (partial) unification framework whose scale is much higher than a TeV. In our view, the LQ explanation for $B$ anomalies could become more convincing if LQs originate from a \textit{most minimal} GUT. Their light masses and specific coupling structures could emerge as a consistency requirement for successful Yukawa unification, so that the benefits of successful unification and explanation of flavor anomalies outweigh the nuisance of additional unprotected scalar masses.

In this work, we indeed demonstrate  that the light scalar particles correcting the $b-\tau$ mass relation could be the same TeV-scale LQs responsible for the $B$ anomalies. We study a minimal SO(10) model, whose Yukawa sector includes merely one scalar multiplet, $126_H$. The SM Higgs doublet and the TeV-scale LQs all live within this $126_H$ representation. This minimal set-up then contains merely one Yukawa coupling matrix, and seemingly cannot reproduce the observed fermion masses and mixing patterns. Yet, the TeV-scale LQs break the desert picture and modify the renormalization group (RG) equations. At the end, we find the masses of the top quark, bottom quark, and $\tau$ lepton to emerge correctly at low energy, although the theory contains just two free parameters --- the Yukawa coupling at the GUT scale, $y_t(M_\text{GUT})$, and the Higgs VEV ratio $\tan\beta$. The LQ-fermion couplings emerge as infrared (IR) fixed points (see Ref.~\cite{Fedele:2023rxb} for a model-independent study of fixed-points), and the correlations among couplings of different LQ types potentially make the  explanation of the $B$ anomalies much less ad-hoc. Although a single scalar multiplet in the Yukawa sector cannot provide flavor mixing angles, we find that in the presence of the TeV-scale LQs, flavor conservation becomes an unstable solution of the RG equations. Flavor mixing can thus serve as an emergent phenomenon when zooming out to larger distance scales. In this work, we do not attempt a global fit to all data, as our result appears to be fairly model-independent and does not rely on the specific choice of the LQ types. A complete explanation of the $B$ anomalies, particularly the new flavor mixing angles, typically requires further modifications and refinements to the light scalar spectrum of $126_H$. Rather than searching for an existence proof, our main goal is to analyze a simple scenario that captures the essential physics and to demonstrate that TeV-scale LQs offer a promising path towards a consistent minimal GUT. Our result makes the LQ explanation to $B$ anomalies more convincing: although fine-tuning scalar LQ masses is still needed, it now arises as a consistency requirement imposed by unification.



The paper is structured as follows. In Section~\ref{minimalSO10}, we revisit the minimal ways to construct the SO(10) Yukawa sector. We explain why $126_H$ is preferred and discuss its shortcomings. Next, we discuss the LQ spectrum in $126_H$, specify the LQ-fermion interactions, and summarize the effective operators relevant for $b\rightarrow c, s$ transitions. In the following subsection, we present an overview on how these LQs can address the $B$ anomalies and examine what additional constrains are imposed by SO(10). In Section~\ref{RGEs}, we analyze how TeV-scale LQs can improve the RG evolution for the Yukawa couplings between light scalars and third-generation fermions. 
We show how the $b-\tau$ mass relationship is improved and how the LQ-fermion couplings exhibit  fixed point behaviors. Since addressing $B$ anomalies requires large flavor mixing angles that are unphysical in SM, we then propose a possible solution to this newly arising problem. In Section~\ref{conclu}, we summarize our main findings and outline the further work needed to strengthen our idea. To demonstrate the model-independence of our results, we show in Appendix~\ref{append} that  $b-\tau$ unification can also be achieved with different light LQ spectra.

\section{Minimal SO(10) GUT}
\label{minimalSO10}

\subsection{The landscape}
We start by revisiting the SO(10) theories with the simplest Yukawa sector~\cite{Mohapatra:1979nn}. As introduced, the spinor representation of SO(10) contains exactly one generation of SM fermions plus a right-handed neutrino:
\begin{equation}
    16_F~=~(Q_L, u_R^c, d_R^c)+(\ell_L, \nu_R^c, e_R^c). 
\end{equation}
As a chiral theory, Fermions in $16_F$ get masses via coupling to the scalars developing non-zero VEVs. An approach is to add a vector $10_H$:
\begin{equation}
    -\mathcal{L}_{Y_{10}}~=~Y_{10}10_H\overline{16_F}16_F^c, \quad 10_H=\Gamma_i\phi_i.
\end{equation}
Here, $\Gamma_i~ (i=1,..10)$ are the Gamma matrices in the SO(10) space. The GUT-scale prediction is robust but a bit boring: degenerate masses for top quark, bottom quark, charged $\tau$ lepton, and Dirac-type $\tau$ neutrino.
This is clearly inconsistent with the low energy data. Alternatively, one can replace the vector $10_H$ with a rank-three tensor $120_H$:
\begin{equation}
    -\mathcal{L}_{Y_{120}}~=~Y_{120}120_H\overline{16_F}16_F^c, \quad 120_H=\Gamma_i\Gamma_j\Gamma_k\phi_{[ijk]}. 
\end{equation}
However, the ten-dimensional Dirac algebra tells that $Y_{120}$ is anti-symmetric in flavor space. $Y_{120}$ vanishes in the single generation limit. In case of three generations, $Y_{120}$ cannot generate heavy masses for the third-generation fermions if the first two generations remain light. This situation is even worse than the one in $10_H$.

For a long time, people believed that the mentioned failure comes from minimal model construction. In our opinion, this trouble somehow stems from the structures of $10_H, 120_H$, which lack  complexity. The last remaining choice, the rank-five (self-dual) tensor $126_H$, improves the situation much. The Yukawa sector now reads: 
\begin{equation}
\label{126H}
    -\mathcal{L}_{Y_{126}}~=~Y_{126}126_H\overline{16_F}16_F^c, \quad 126_H=\Gamma_i\Gamma_j\Gamma_k\Gamma_l\Gamma_m\phi_{[ijklm]}. 
\end{equation}
Firstly, $126_H$ is tailor-made for tiny neutrino masses. $126_H$ contains a vacuum singlet under $G_{\text{SM}}$ and can give the right-handed neutrino $\nu_R$ a high-scale mass term. Upon integrating out $\nu_R$, the dim-5 Weinberg operator~\cite{Weinberg:1979sa} can generate a tiny mass for the active neutrino living in $\ell_L$~\cite{Minkowski:1977sc, Mohapatra:1979ia}. In addition, $126_H$ contains two Higgs doublets with opposite $SU(2)_R$ isospin. As a result, one of the Higgses only couples to $\overline{Q_L}u_R, \overline{L_L}\nu_R$ and the other one only couples to $\overline{Q_L}b_R, \overline{Q_L}e_R$. A hierarchical ratio $m_t/m_b$ is now possible, as long as the ratio $\tan\beta$ of the two Higgs doublets VEVs  is large. Furthermore, $Y_{126}$ is symmetric and can be assigned third-generation specific, which gives the desired hierarchy structure among the charged fermion masses of the three generations. Indeed, Eq.~(\ref{126H}) cannot account for quark mixing and is unlikely to yield the precise quantities for the first- and second-generation fermion masses. However, we do not think these shortcomings require extending the Yukawa sector immediately. If one neglects Yukawa couplings much smaller than unity, the light-flavor structure contains merely vanishing or unphysical observables. Large neutrino mixing is not problematic either, because the Yukawa couplings enter the left-handed neutrino mass matrix $M_{\nu_L}$ non-linearly. The relation $M_{\nu_L}\propto Y_{126}$ could break down even with higher-order corrections. Therefore, the $\mathcal{O}(1)$ predictions of $126_H$ are approximately correct. In the following, we refer to the theory whose Yukawa sector can be well-approximated by Eq.~(\ref{126H}) as `minimal SO(10) GUT'.

\begin{figure}[t]
    \centering
    \includegraphics[width=0.7\linewidth]{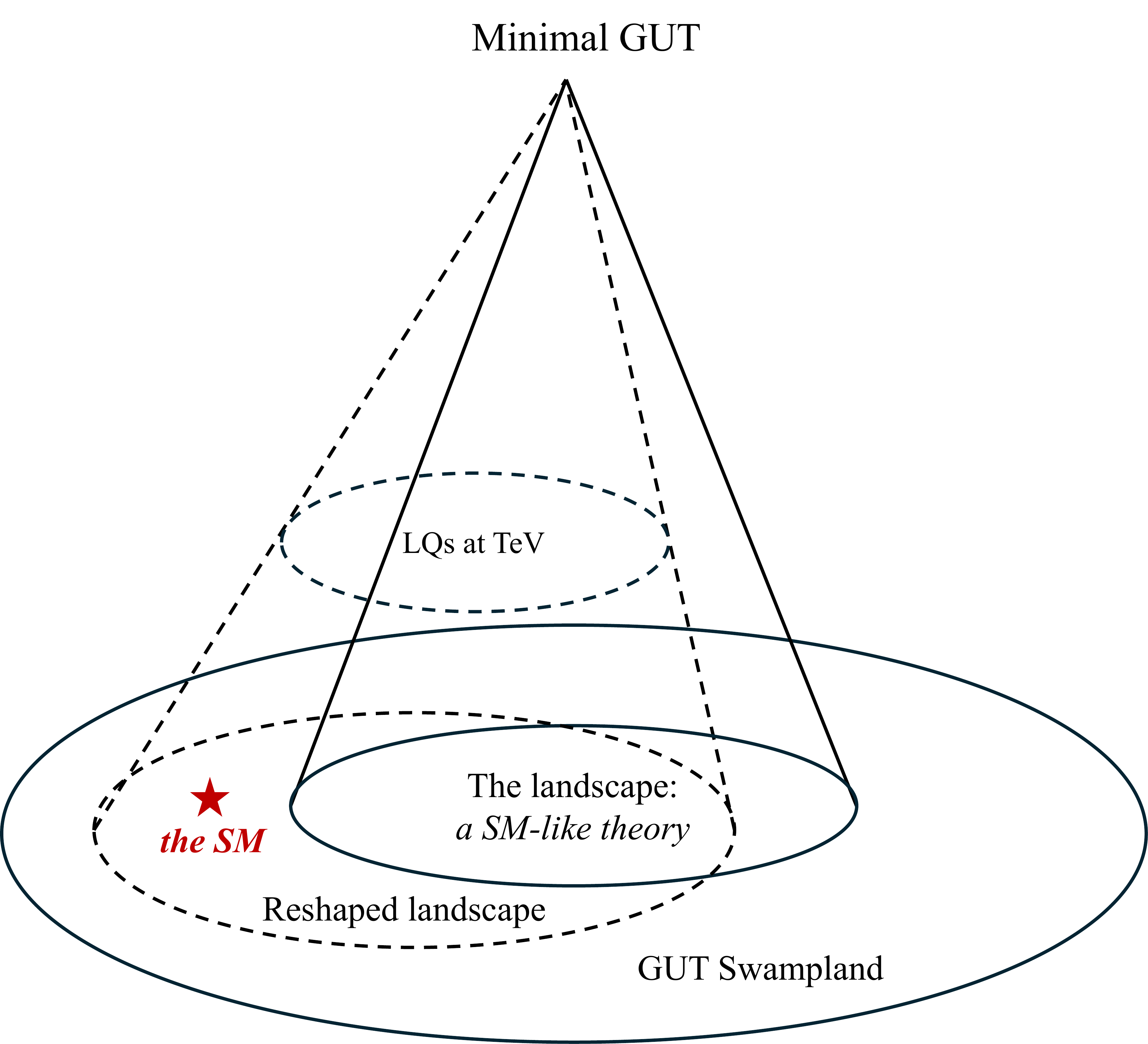}
    \caption{The small solid ellipse represents the landscape of minimal SO(10), which is \textit{a SM-like theory}. The red star marks \textit{the SM} itself. It is not consistent with the minimal theory and therefore lies in the GUT swampland (shown with the large solid ellipse). The dashed ellipse illustrates the reshaped landscape induced by TeV-scale LQs, within which \textit{the SM} is now included.} 
    \label{landscape}
\end{figure}

However, the $b-\tau$ mass relation remains an issue: $126_H$ predicts $m_{\tau}=3m_b$ at $M_{\text{GUT}}$~\cite{Mohapatra:1979nn}, but the experimental measurements, combined with the SM RG equations, indicate $m_{\tau}= 1.67m_b$ at $M_{\text{GUT}}$~\cite{Martin:2019lqd,  Huang:2020hdv}. This discrepancy cannot be ignored. Therefore, we have to conclude that the low-energy theory of minimal SO(10) contains all SM fields with couplings close to --- but still not reasonably agreeing with --- the measured quantities. The landscape of this minimal theory is \textit{a SM-like theory}, while \textit{the SM} itself lies in the swampland, as illustrated in Figure~\ref{landscape}. Can \textit{the SM} be included inside the landscape without modifying the deep-UV physics? Our conjecture is: if some scalar particles living in $126_H$, such as LQs, are fine-tuned to be light, they can change the RG equation and reshape the boundary of the minimal SO(10) landscape, which may include \textit{the SM}.

\subsection{Leptoquarks in $126_H$}
To discuss the LQ spectrum, we decompose the spectrum of $126_H$ with the Pati-Salam type subgroup $SU(4)_c\times SU(2)_L\times SU(2)_R$ by~\cite{Slansky:1981yr}: 
\begin{equation}
    126_H~=~(6_c,1_L,1_R)~+~(10_c, 3_L, 1_R)~+~(\overline{10_c}, 1_L,3_R)~+~(15_c,2_L,2_R). 
\end{equation}
The $SU(4)_c$ group takes the anomaly-free $U(1)_{B-L}$ charge as the `fourth color', implying that  leptons and quarks can convert into each other. Here, the scalar LQs are identified as the color triplet mediators:
\begin{equation}
    \begin{aligned}
        (6_c,1_L,1_R)~&\supset~S_1(3,1,-1/3)+S_1'(\overline{3},1,1/3),\\
        (10_c, 3_L, 1_R)~&\supset~S_3(3,3,-1/3),\\
        (\overline{10_c}, 1_L,3_R)~&\supset~\bar{S}_1(\overline{3},1,-2/3)
        +S_1''(\overline{3},1,1/3)+\widetilde{S}_1(\overline{3},1,4/3),\\
        (15_c,2_L,2_R)~&\supset~ R_2(\overline{3},2,-7/6)
        +\widetilde{R}_2(\overline{3},2,-1/6)+R_2'(3,2,7/6)+\widetilde{R}_2'(3,2,1/6).\\
    \end{aligned}
\end{equation}
We follow the LQ notation of Ref.~\cite{Dorsner:2016wpm} up to charge conjugation, with the numbers in the parentheses indicating their representations under $G_{\text{SM}}$. $\bar{S}_1$, $S_1''$, and $ \widetilde{S}_1$ form an $SU(2)_R$ triplet, while $S_1'$ is an $SU(2)_R$ singlet. They carry symmetric and antisymmetric $SU(2)_R$ indices, respectively, and can be written in a compact form as: 
\begin{equation}
    \widehat{S}_3~=~
    \left(
        \begin{array}{cc}
           \overline{S}_1  & S''_1/\sqrt{2} \\
            S''_1/\sqrt{2} & \widetilde{S}_1
        \end{array}
        \right)
    \qquad
    \widehat{S}_1~=~
    \left(
        \begin{array}{cc}
           0  & S'_1/\sqrt{2} \\
            -S'_1/{\sqrt{2}} & 0
        \end{array}
        \right)
\end{equation}
Both $\widehat{S}_3$ and $\widehat{S}_1$ have $Q_{B-L}=1/3$. The pairs $(R_2, \widetilde{R}_2)$ and $(R_2', \widetilde{R}_2')$ are two distinct $SU(2)_R$ doublets, with $Q_{B-L}= 2/3$ and $Q_{B-L}=-2/3$ respectively. Their Yukawa interactions with fermions are given by: 
\begin{equation}
\label{chirality}
    \begin{aligned}
        -\mathcal{L}_Y^{\text{LQ}}~=~&Y_3^{LL}\overline{Q_L}S_3L_L^c+
        Y_3^{RR}\overline{Q_R^c}
        \widehat{S}_3
        L_R+
        Y_1^{LL}\overline{Q_L}S_1L_L^c+Y_1^{RR}\overline{Q_R^c}\widehat{S}_1L_R\\
        &+Y_2^{LR}\overline{Q_L}(R_2', \widetilde{R}_2')L_R
        +Y_2^{RL}\overline{Q_R^c}(R_2, \widetilde{R}_2)L_L^c+\text{h.c.}
    \end{aligned}
\end{equation}
Here,$^c$ denotes the standard charge conjugation operator, $Q_R=(u_R, d_R),~L_R=(\ell_R, \nu_R)$ are $SU(2)_R$ doublets, and flavor indices are implicit. Eq.~(\ref{chirality}) shows $R$-type LQs couple to quarks and leptons with opposite chiralities, while the $S$-type LQs --- $S_3, S_1$ for left-handed fields and $\widehat{S}_3, \widehat{S}_1$ for right-handed fields --- are chirality-specific.

The Yukawa couplings $Y_{1,2,3}$ are $3\times3$ matrices in flavor space. At the GUT scale, they are aligned with the quark mass matrices and therefore inherit an approximate $U(2)$ structure: 
\begin{equation}
\label{FlavorUV}
    Y_{1,2,3}~\propto~Y_{126}~\sim~ \begin{pmatrix} ~\epsilon~ & ~0~ & ~0~ \cr ~0~ & ~\epsilon'~ & ~0~ \cr ~0~ & ~0~ & ~1~ \end{pmatrix}, \quad \text{at GUT scale.}
\end{equation}
Here, $\epsilon\ll\epsilon'\ll1$, following the known hierarchy of quark masses. 
This structure, if well preserved at low energies, yields the approximate $U(2)$ symmetry needed to suppress the processes such as $K_L\rightarrow\mu e$. In this unification framework, the $U(2)$ LQ-fermions coupling structures are no longer ad-hoc, but emerge naturally since they have the same origin as quark masses. In the following discussion, we focus on third-generation specific LQ couplings and include the second generation only when they are needed for the flavor-violating transitions.

\begin{table}[t]
    \centering
    \renewcommand\arraystretch{2}
    \begin{tabular}{c|c c c }
    \hline
    ~ & $S_3$ & $S_1$ & $\widetilde{S}_1$  \\
    \hline
        $b\rightarrow c \tau \nu$ & $(\overline{c}_L\gamma^{\mu}b_L)(\overline{\tau}_L\gamma_{\mu}\nu_L)$  &\renewcommand\arraystretch{1.5}
            \begin{tabular}{c}
                 $(\overline{c}_R \sigma^{\mu\nu} b_L)(\overline{\tau}_R\sigma_{\mu\nu} \nu_L)$  \\
                 $(\overline{c}_L\gamma^{\mu}b_L)(\overline{\tau}_L\gamma_{\mu}\nu_L)$\\
                 $(\overline{c_R}b_L)(\overline{\tau}_R \nu_L)$ 
            \end{tabular}  & $-$   \\
       $ b\rightarrow s \tau\tau$ & $(\overline{s}_L\gamma^{\mu}b_L)(\overline{\tau}_L\gamma_{\mu}\tau_L)$ &$-$&
       $(\overline{s}_R\gamma^{\mu}b_R)(\overline{\tau}_R\gamma_{\mu}\tau_R)$ \\
       $ b\rightarrow s  \nu\nu$  & $(\overline{s}_L\gamma^{\mu}b_L)(\overline{\nu}_L\gamma_{\mu}\nu_L)$ & $(\overline{s}_L\gamma^{\mu}b_L)(\overline{\nu}_L\gamma_{\mu}\nu_L)$ & $-$ \\
       \hline
       \hline
    ~ & $R_2$ & $\widetilde{R}_2$ \\
    \hline
        $b\rightarrow c \tau \nu$ & \renewcommand\arraystretch{1.5}
            \begin{tabular}{c}
                 $(\overline{c}_R \sigma^{\mu\nu} b_L)(\overline{\tau}_R\sigma_{\mu\nu} \nu_L)$  \\
                 $(\overline{c_R}b_L)(\overline{\tau}_R \nu_L)$ 
            \end{tabular}   & $-$
                   \\
       $ b\rightarrow s \tau\tau$ &  $(\overline{s}_L\gamma^{\mu}b_L)(\overline{\tau}_R\gamma_{\mu}\tau_R)$ & $(\overline{s}_R\gamma^{\mu}b_R)(\overline{\tau}_L\gamma_{\mu}\tau_L)$ \\
       $ b\rightarrow s  \nu\nu$    & $-$ & $(\overline{s}_R\gamma^{\mu}b_R)(\overline{\nu}_L\gamma_{\mu}\nu_L)$ \\
       \hline
    \end{tabular}
    \caption{Effective operators induced by LQ exchange, expressed in the standard basis. We assume maximal mixing among the LQs with same quantum numbers under $G_{\text{SM}}$. Top quarks and right-handed neutrinos are omitted. Except for charm and strange quarks, we do not include other second- and first-generation fermions either.} 
    \label{eftS}
\end{table}

In the energy regime far below 1 TeV, LQs in Eq.~(\ref{chirality}) can be safely integrated out, and then the scalar-type effective operators with the form $(\overline{q}\ell)(\overline{\ell'}q')$ are generated at tree-level. We apply a Fierz transformation to convert these operators into the standard basis, as summarized in Table~\ref{eftS}. These operators follow the general discussion shown in Ref.~\cite{Dorsner:2016wpm} and can also be directly inferred from the chirality structure of the LQ-fermion couplings. It is worth noting that the left-right symmetry is not manifest in the table, because $\nu_R$ and the top quark are not included. Moreover, the scalar- and tensor-type operators require sizable $R_2-(R_2')^c$ and $S_1-(S_1')^c$ (or $S_1-(S_1'')^c$) mixing. Without such mixings, a single LQ in the generic basis cannot couple simultaneously to both $Q_L$ and $Q_R$, and thus leads to merely the vector-type operators.

\subsection{Addressing the $B$ anomalies}
\label{anomalies}

TeV scale LQs are related to the long-standing anomalies observed in semi-leptonic B decays. The $S_3$ LQ contributes to all the $b\rightarrow c, s$ transitions with SM-like operators. If $S_3$ is the unique LQ responsible for the $R(D^{(*)})$ anomaly, $C_L^{\nu}$, the coefficient of $(\overline{s}_L\gamma^{\mu}b_L)(\overline{\nu}_L\gamma^{\mu}\nu_L)$ is  fixed by $SU(2)_L$ invariance. The predicted value for $C_L^{\nu}$ is too large to be compatible with the current bounds on $b\rightarrow s\overline{\nu}\nu$ transitions. A simple way out is to combine $S_3$ with $S_1$, whose contributions to $C_L^{\nu}$ can partly cancel~\cite{Crivellin:2017zlb, Crivellin:2019dwb} the one of the former LQ. Assuming a cancellation of about $60\%$ in $C_L^{\nu}$, and including the operators induced by $S_1$, $R(D^{(*)})$ can be consistently explained at $1\sigma$ level~\cite{Crivellin:2025qsq}. The constraints on $C_L^{\nu}$ can be relaxed since the recent Belle-II data with the inclusive tagging method indicates $\text{Br}(B\rightarrow K\overline{\nu}\nu)$ exceeds the SM prediction by a factor of 5.4. Yet $\text{Br}(B\rightarrow K^*\overline{\nu}\nu)$ still imposes a tight bound and necessitates the right-handed operator $(\overline{s}_R\gamma^{\mu}b_R)(\overline{\nu}_L\gamma^{\mu}\nu_L)$ generated by $\widetilde{R}_2$. If its coefficient $C_R^{\nu}$ takes a proper value, it can suppress the $B\rightarrow K^*$ amplitude while allowing a sizable $B\rightarrow K$ rate~\cite{Bause:2023mfe, He:2023bnk}. $SU(2)_L$ invariance also implies $b\rightarrow s\overline{\tau}\tau$ transitions. The current limit on $\text{Br}(B\rightarrow K\overline{\tau}\tau)$ is weak due the experimental difficulty in identifying $\tau$ leptons. Interestingly, if closing the $\tau$ loop and attaching it to an off-shell photon, the penguin diagram can induce $b\rightarrow s\ell\ell$ ($\ell=e,\mu$) via lepton flavor universal operators~\cite{Crivellin:2018yvo}. The coefficient $C_9^U\sim-1$ can account for the $B\rightarrow K\ell\ell$ anomalies without violating the bound from $R(K^{(*)})$~\cite{Alguero:2023jeh}. Although $B_s-\bar{B}_s$ mixing constraints disfavor the best-fit value for $C_9^U$, moderate cancellation by additional operators containing right-handed quarks can relax this tension~\cite{Crivellin:2015era, Capdevila:2023yhq}. Such operators can be generated by $\widetilde{S}_1$ or $\widetilde{R}_2$.

The $R_2$ LQ can also explain the $R(D^{(*)})$ anomaly with the $(\overline{c}_R \sigma^{\nu\nu} b_L)(\overline{\tau}_R\sigma_{\nu\nu} \nu_L)$ and $(\overline{c_R}b_L)(\overline{\tau}_R \nu_L)$ operators~\cite{Sakaki:2013bfa, Becirevic:2018afm, Cheung:2020sbq}, and it does not lead to the $b\rightarrow s \overline{\nu}\nu$ transition. This solution requires an $\mathcal{O}(1)$ $R_2-(R_2')^c$ mixing angle, which brings an inhomogeneous term to the RG equation for the $\tau$ lepton mass~\cite{Fedele:2023rxb}:
\begin{equation}
   16\pi^2  \frac{d }{ d \ln\mu}m_{\tau} ~=~...-6 m_t (Y_2^{LR})_{33} (Y_2^{RL})_{33}. 
\end{equation}
The $(3,3)$ elements of $Y_2^{LR}$ and $Y_2^{RL}$ are both $\mathcal{O}(1)$ and therefore can lead to an unsuppressed additive correction to the $\tau$ lepton mass. This situation is somehow similar to the SUSY threshold correction to the bottom quark mass in the large $\tan\beta$ regime~\cite{Hall:1993gn,Carena:1993ag,Carena:1999py}. Here, the physical correction to $m_{\tau}$ is further enhanced by a logarithmic factor $\log(M_{\text{GUT}}/m_{R_2})\sim 30$. The predicted $m_{\tau}$ is then too large and cannot be consistent with the observed value in a minimal theory. Moreover, $R_2$ carries a large hyper-charge $7/6$ and significantly accelerates the running of $g_1$, the $U(1)_Y$ coupling. With the RG equations of SM, $g_1$ already increases rapidly and meets the other gauge couplings at around $10^{13}$ GeV~\cite{Amaldi:1991cn, Langacker:1991an}, a scale too low to satisfy proton decay constrains. Light $R_2$ would worsen this problem~\cite{Dorsner:2006dj, Dorsner:2007fy}, making gauge coupling unification even more challenging.

The $S_1$ LQ is constrained to be as heavy as the GUT scale, because it always couples to a pair of quarks in the minimal GUT framework~\cite{Patel:2022wya} and induces proton decay. On the other hand, the diquark couplings of $S_3, \overline{S}_1, \widetilde{S}_1$ are absent at tree level. This absence is not accidental but a consequence of the $U(1)_{\text{PQ}}$ symmetry contained in Eq.~(\ref{126H}).
\begin{equation}
    16_F\rightarrow 16_F e^{i\theta}, \quad 126_H\rightarrow 126_He^{2i\theta}. 
\end{equation}
As a phase rotation of complex fields, this symmetry also emerges in the gauge sector. Taking $S_3$ as an example, although the SM gauge symmetry allows it to couple to both $(\overline{Q}_LL_L^c)$ and $(\overline{Q_L^c}Q_L)$, the minimal Yukawa sector does not contain the $\overline{Q_L^c}S_3Q_L$ term because it is not invariant under $U(1)_{\text{PQ}}$. To suppress these diquark couplings also at the loop level, the PQ symmetry should also be well preserved in the scalar potential. The $\eta_2(126_H)^4$ and $\gamma_2 (45_H)^2(126_H)^2$ terms~\cite{Bertolini:2012im} explicitly break the PQ symmetry. To ensure that the proton decay amplitude is suppressed by $M_{\text{GUT}}^2$, the magnitudes of $\gamma_2$ and $\eta_2$ can be much larger than about $\mathcal{O}\left(\frac{m_{S}}{M_{\text{GUT}}}\right)^2$, where the $m_S$ denotes lightest mass among $S_3, \overline{S}_1$ and $\widetilde{S}_1$.\footnote{Since the hierarchy could be large, we omit the possible loop factors, which can relax the bound by $(16\pi^2)^n$, where $n$ is the loop order.} If $m_S\sim$ TeV, $\gamma_2$ and $\eta_2$ are constrained extremely small. By itself, this does not directly introduce a new hierarchy puzzle, since $\gamma_2=\eta_2=0$ enhances the PQ symmetry. However, the $H_u-H_d$ mixing term is proportional to $\gamma_2 M_{\text{GUT}}^2$. To achieve a sizable mixing angle $\beta$, the mass of $H_d$ cannot be far larger than $\sqrt{\gamma_2} M_{\text{GUT}}\lesssim \mathcal{O}(m_{\text{S}})$. This requires additional fine-tuning, while it enriches the low-energy spectrum with an extra Higgs doublet. Its charged component can contribute to $b\rightarrow c\tau\nu$~\cite{Crivellin:2012ye, Crivellin:2013wna, Iguro:2022uzz, Blanke:2022pjy} and further relax the tension between explaining $R(D^{(*)})$ and avoiding $B\rightarrow K \overline{\nu}\nu$ constrains.

To conclude, $126_H$ contains six types of LQs: $S_3, S_1, \overline{S}_1, \widetilde{S}_1$ and  $R_2, \widetilde{R}_2$. Among them, $S_1$ is excluded from being light due to the proton decay constrains. $R_2$ is disfavored by both Yukawa and gauge coupling unification. $\overline{S}_1$ only couples to $\nu_R$. On the other hand, $S_3, \widetilde{S}_1$ and $\widetilde{R}_2$ can address the $B$ anomalies, and a consistent explanation typically requires a combination of two or all of them. 
It is worth to remark here that the precise flavor data is not strongly correlated to Yukawa unification. The coefficients for the effective operators addressing the B anomalies are suppressed by $1/M_{\text{LQ}}^2$. Here, $M_{\text{LQ}}$ represents the mass of the relevant LQ. Varying $M_{\text{LQ}}$ around the TeV scale (to fit the B anomalies) only brings sub-leading noise to the $b-\tau$ mass relationship, because the effects of RG evolution is proportional to $\log(M_{\text{LQ}}/M_{\text{GUT}})$. Therefore, we do not examine in detail  the regions preferred by the precise flavor data, but only stress the B anomalies are data-driven motivations for light LQs.


\section{Improved RG evolution}
\label{RGEs}

\subsection{$b-\tau$ masses}
To quantify the modified $b-\tau$ mass relation, we first specify the relevant energy scales. Proton decay constraints and hints from gauge coupling unification imply that the breaking scale of SO(10) generally lies around $10^{16}$ GeV~\cite{Deshpande:1992au, Deshpande:1992em, Bertolini:2009qj, Ohlsson:2019sja}. The intermediate scale can be defined by the heaviest right-handed neutrino mass $m_{\nu_R}$, satisfying the relation:
\begin{equation}
    m_{\nu_L}~=~\frac{m_{\nu_D}^2}{m_{\nu_R}}~=~\frac{(3 m_t)^2}{m_{\nu_R}}~<~ 0.064~\text{eV~\cite{DESI:2025zgx}}~(0.45~\text{eV~\cite{KATRIN:2024cdt}}). 
\end{equation}
Considering the cosmological constraint from the combination of DESI and CMB~\cite{DESI:2025zgx}, $m_{\nu_R}$ is larger than around $10^{15}$ GeV. The limit from KATRIN~\cite{KATRIN:2024cdt} is weaker by a factor of about 7 but it is model-independent. In either case, the intermediate scale $m_{\nu_R}$ remains high\footnote{This only holds with the non-minimal Yukawa sector.} and fairly close to $10^{16}$ GeV. The low energy spectrum contains SM particles and the TeV-scale scalar particles originating from $126_H$, including an additional Higgs doublet as well as $S_3, \widetilde{S}_1, \widetilde{R}_2$. We consider a simpler scenario without $\widetilde{S}_1$ as well, although it may not achieve a global fit as good as the full three-LQ case. For completeness, we also analyze the $S_3, \widetilde{S}_1$ scenario to highlight model-independence. Since the precise mass values are not the focus of this work, we simply assume a common mass of 1 TeV for all light LQs and the Higgs doublet, and set the masses of all other BSM particles to $M_{\text{GUT}}\equiv10^{16}$ GeV. This two-scale approximation is valid at the leading-log order. A more specified mass distribution cannot introduce large hierarchies which would  significantly change the RG evolution.

The Yukawa couplings relevant for the running are defined by: 
\begin{equation}
\label{yukawaDef}
\begin{aligned}
    -\mathcal{L}_Y~=&~y_t \overline{Q_L^3}t_R H_u+ y_b \overline{Q_L^3}b_R H_d+y_{\tau} \overline{L_L^3}\tau_R H_d\\
    &~+y_1 \overline{b_R^c}\tau_R \widetilde{S}_1+y_2 \overline{b_R^c} L_L^{3c} \widetilde{R}_2+ y_3 \overline{Q_L^{3c}}L_L^3 S_3+\text{h.c.}
\end{aligned}
\end{equation}
Here, $Q_L^3=(t_L, b_L)$ and $L_L^3=(\tau_L, \nu_{\tau L})$ are third-generation specific, as we do not include light flavors at this stage. All these Yukawa couplings can be chosen real. At $M_{\text{GUT}}$, they are related by the group structure of $126_H$:
\begin{equation}
\label{initial}
    y_b=y_t, \quad y_{\tau}=-3 y_{b}, \quad y_1=y_2=y_3=2\sqrt{3} y_t, \quad \text{at GUT scale. }
\end{equation}
These relations serve as the initial condition of RG evolution from IR to UV. The new factor $2\sqrt{3}$ can be interpreted as a generalized Clebsch-Gordan coefficient, and we take this number from Ref.~\cite{Patel:2022wya}. This factor enhances the LQ Yukawa coupling over the top quark one by more than a factor of three, yielding a large impact on the running.

The explicit RG equations are form Refs.~\cite{Fedele:2023rxb, Grzadkowski:1987wr, Branco:2011iw}. We reduce them by keeping only the third-generation related couplings:
\begin{equation}
\label{runningEq}
    \begin{aligned}
       16\pi^2 \frac{d}{d \ln{\mu}}y_t~=&~y_t\left( -\frac{17 g_1^2}{12}-\frac{9 g_2^2}{4}-8 g_3^2+\frac{9 y_t^2}{2}+\frac{y_b^2}{2}+\frac{3 y_3^2}{2}\right), \\
16\pi^2\frac{d}{d \ln{\mu}}y_b~=&~y_b\left( -\frac{5 g_1^2}{12}-\frac{9 g_2^2}{4}-8 g_3^2+\frac{y_t^2}{2}+\frac{9 y_b^2}{2}+y_{\tau
   }^2+\frac{y_1^2}{2}+y_2^2+\frac{3 y_3^2}{2}\right), \\
16\pi^2\frac{d}{d \ln{\mu}}y_{\tau}~=&~y_{\tau}\left( -\frac{15 g_1^2}{4}-\frac{9 g_2^2}{4}+\frac{5 y_{\tau }^2}{2}+3 y_b^2+\frac{3 y_1^2}{2}+\frac{3 y_2^2}{2}+\frac{9
   y_3^2}{2}\right), \\
16\pi^2\frac{d}{d \ln{\mu}}y_1~=&~y_1\left(-2 g_1^2-4 g_3^2+ y_b^2+\frac{y_{\tau }^2}{2}+3 y_1^2+y_2^2\right), \\
16\pi^2\frac{d}{d \ln{\mu}}y_2~=&~ y_2\left(-\frac{13 g_1^2}{20}-\frac{9 g_2^2}{4}-4 g_3^2+y_b^2+\frac{y_{\tau }^2}{2}+\frac{y_1^2}{2}+\frac{7
   y_2^2}{2}+\frac{9 y_3^2}{2}\right), \\
16\pi^2\frac{d}{d \ln{\mu}}y_3~=&~y_3\left( -\frac{g_1^2}{2}-\frac{9 g_2^2}{2}-4 g_3^2+\frac{y_t^2}{2}+\frac{y_b^2}{2}+\frac{y_{\tau }^2}{2}+\frac{3
   y_2^2}{2}+8 y_3^2 \right).\\
    \end{aligned}
\end{equation}
Here, $g_1,g_2,$ and $g_3$ are the coupling strengths of $U(1)_Y, SU(2)_L,$ and $SU(3)_c$, respectively. No inhomogeneous terms arise because the $U(1)_{\text{PQ}}$ symmetry ensures each LQ only couples to a single type of fermion bilinear. This feature further simplifies the RG equation compared to the general case~\cite{Fedele:2023rxb}. The beta function for every coupling is always proportional to itself. Moreover, the running receives negative contributions from the gauge interactions and positive contributions from Yukawa couplings.

The IR observables are the third-generation charged fermions' masses at 1 TeV, defined by: 
\begin{equation}
\label{match}
    m_t=\frac{1}{\sqrt{2}}y_t v\sin\beta  , \quad m_b=\frac{1}{\sqrt2}y_b v\cos\beta, \quad m_{\tau}=\frac{1}{\sqrt{2}}y_{\tau}v\cos\beta, \quad v\equiv 246 ~\text{GeV}. 
\end{equation}
Their specific values at 1 TeV can be found in Ref.~\cite{Antusch:2013jca}
\begin{equation}
\label{mTeV}
    m_t=151.1\pm1.6~\text{GeV}, \quad m_b=2.414\pm 0.024~\text{GeV}, \quad m_{\tau}~=~1.7780\pm 0.0014~\text{GeV}. 
\end{equation}
As a result, there are merely two free input parameters in the system: the GUT-scale value of $y_t$ and the Higgs VEV ratio $\cot\beta$. They are constrained by $m_t$ and $m_b$. The $\tau$ lepton mass then becomes a prediction and thus must reasonably agree with the experimental value.

\begin{figure}[t!]
    \centering
    \includegraphics[width=0.49\linewidth]{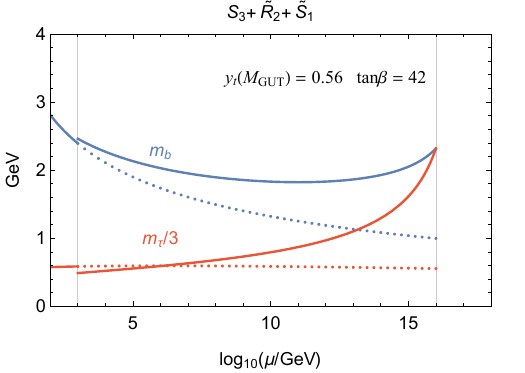}
    \includegraphics[width=0.5\linewidth]{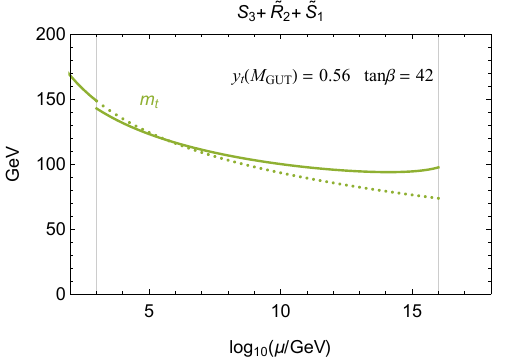}
    \caption{RG evolution of the third-generation charged fermions masses from $10^2$ to $10^{16}$ GeV. The solid (dotted) lines indicate the scenario with (without) LQs at TeV scale. 
    We fix $y_t(M_{\text{GUT}})=0.56$ to get the correct value of $m_t$ and Yukawa unification implies $\tan\beta=42$. The gray vertical lines indicate $M_{\text{GUT}}$ and the light LQ threshold.}
    \label{S3R2}
\end{figure}

We illustrate how light LQs from $126_H$ change the mass running in Figure~\ref{S3R2}. We assume exact Yukawa unification at $M_{\text{GUT}}$ (Eq.~(\ref{initial}) holds exactly) and fix the two input parameters to $y_t(M_\text{GUT})=0.56$ and $\tan\beta=42$. The solid lines indicate how the $b, \tau, t$ masses evolve from $10^{16}$ GeV down to $10^2$ GeV. For $1~\text{TeV}<\mu<10^{16}~\text{GeV}$, we use the running equation shown in Eq.~(\ref{runningEq}). Below $1~\text{TeV}$, we directly interpolate the SM running data included in the SMDR package.~\cite{Martin:2019lqd}. The resulting $b, \tau, t$ masses at 1 TeV can agree reasonably well with the values extracted from the SM. While the matching at 1 TeV would be perfect if the solid lines were continuous, the gaps visible in the plots are not problematic. At leading-log level, TeV-scale LQs can reduce the $b-\tau$ mass tension from an $\mathcal{O}(1)$ discrepancy to a suppressed $\mathcal{O}(\epsilon)$ mismatch. For comparison, we also show the evolution without TeV-scale LQs, by plotting the SM running parameters from Ref.~\cite{Martin:2019lqd} with dots for $\mu>1$ TeV. The red and blue dotted curves still diverge significantly at $10^{16}$ GeV, indicating the relation $m_b= m_{\tau}/3$ cannot be reached if the theory below $M_{\text{GUT}}$ is SM alone.


We also explore scenarios with only $S_3+\widetilde{R}_2$ and only $S_3+\widetilde{S}_1$. 
For both scenarios we are able to find values for $y_t(M_\text{GUT})$ and $\tan\beta$ leading to
successful Yukawa unification at $10^{16}$ GeV and reasonably good matching relationships at 1 TeV. Related plots are shown in Appendix~\ref{append}. This universal feature suggests that the improved $b-\tau$ unification is not an accidental outcome of a particular LQ choice, but rather a model-independent effect from colored scalar fields. We understand the underlying reason as follows. If the IR theory is simply the SM, $(m_{\tau}/3)$ is too small and remains nearly a constant. $m_b$ decreases as the energy scale $\mu$ goes up, mainly due to QCD effects, but not fast enough to meet $(m_{\tau}/3)$ at $M_{\text{GUT}}$. As shown in Eq.~(\ref{runningEq}), the additional homogeneous terms in the $\beta$-function of $y_{\tau}$ are always positive and rapidly drive $y_{\tau}$ towards a Landau-pole at UV. Although a similar behavior is also seen for $y_{b}$, its growth is slower because $y_{\tau}$ receives an extra enhancement by a factor of $N_c=3$.

\subsection{LQ-fermion couplings}

\begin{figure}[t!]
    \centering
    \includegraphics[width=0.492\linewidth]{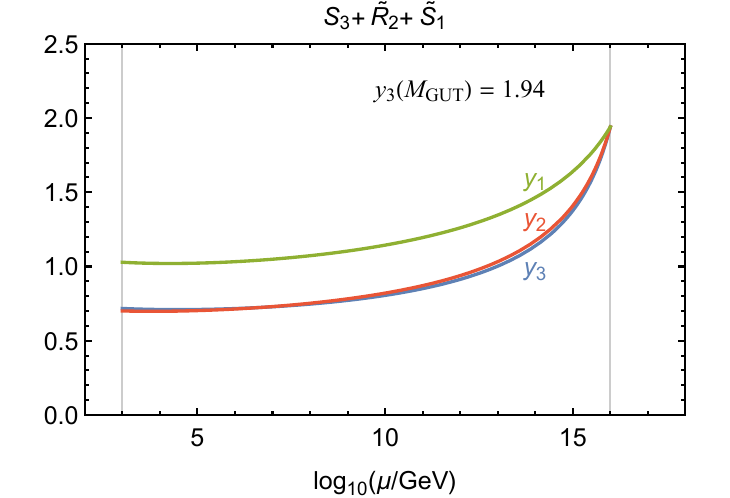}
    \includegraphics[width=0.492\linewidth]{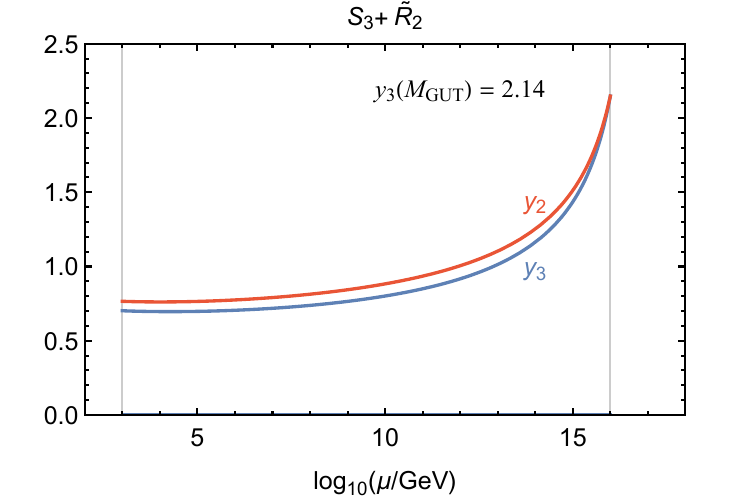}
    \caption{RG evolution of the LQ-fermion coupling $y_1,y_2$ and $y_3$ (green, red, and blue,  respectively) from $10^2$ to $10^{16}$ GeV. The right panel shows the scenario without $\widetilde{S}_1$ in the light spectrum, with $y_t(M_{\text{GUT}} )=0.62$. All other parameters are as in Figure~\ref{S3R2}.}
    \label{LQyuk}
\end{figure}

The LQ-fermion couplings $y_1, y_2, y_3$ are also outcomes from the RG evolution. They are correlated at $M_{\text{GUT}}$ by Eq.~(\ref{initial}) and evolve together with $y_t, y_b, y_{\tau}$ according to Eq.~(\ref{runningEq}). Since $y_t(M_{\text{GUT}})$ is fixed by the measured value of $m_t$, $y_1, y_2$ and $y_3$ in the range $1~\text{TeV}<\mu<M_{\text{GUT}}$ become predictions of the minimal GUT, as shown in the left panel of Figure~\ref{LQyuk}. A fixed-point behavior~\cite{Pendleton:1980as, Hill:1980sq} emerges: all LQ-fermion couplings drop rapidly near $M_{\text{GUT}}$ and remain nearly  constant for $\mu\lesssim 10^5\sim 10^{10}$ GeV. This behavior is similar to what recently found in Ref.~\cite{Fedele:2023rxb}, that the LQ-fermions couplings approach IR fixed-points of $0.5\sim1.0$. Interestingly, the LQ-type considered here is different, suggesting that this feature is a rather general consequence from RG evolution.

Eq.~(\ref{runningEq}) contains no inhomogeneous terms, so the IR fixed points can be solved analytically. Neglecting $g_1, g_2$ for simplicity and assuming $g_3$ varies sufficiently slowly, we find the fixed point solution: 
\begin{equation}
    y_1~=~0.88g_3, \quad y_2~=~0.55g_3, \quad y_3=0.51 g_3, \quad \text{at deep IR.}
\end{equation}
As a result, LQ-fermion couplings are numerically close to $g_3$ and $y_t$ at TeV-scale. Together with Ref.~\cite{Fedele:2023rxb}, this supports the scalar LQ explanation of the $B$ anomalies, because $\mathcal{O}(1)$ couplings are typically required to account for the data. Moreover, we find $y_2$ and $y_3$ almost coincide at all scales. Although this is an accidental result, it can be understood by setting $y_2=y_3$ and neglecting $g_1, g_2, y_t, y_b$ in Eq.~(\ref{runningEq}):
\begin{equation}
    \label{runningOverlap}
\begin{aligned}
    16\pi^2\frac{d \ln{y_2}}{d \ln{\mu}}~=&~ \left(-4 g_3^2+\frac{y_{\tau }^2}{2}+\frac{19 y_3^2}{2}\left(1+\frac{y_1^2-3y_3^2}{19y_3^2}\right)\right), \\
16\pi^2\frac{d \ln{y_3}}{d \ln{\mu}}~=&~\left(-4 g_3^2+\frac{y_{\tau }^2}{2}+\frac{19
   y_3^2}{2} \right).\\
\end{aligned}
\end{equation}
This implies $\frac{d \ln{y_2}}{d \ln{\mu}}\approx \frac{d \ln{y_3}}{d \ln{\mu}}$ so that the ratio $\frac{y_2}{y_3}$ remains nearly a constant under the evolution. A similar result is observed when $\widetilde{S}_1$ is not in the light spectrum ($y_1=0$), as shown in the right panel of Figure~\ref{LQyuk}. We consider the unification prediction $y_2=y_3$ as a critical ingredient for  a consistent explanation of the $B$ anomalies. As explained in subsection~\ref{anomalies}, the $S_3$ and $\widetilde{R}_2$ contributions to $B\rightarrow K^{*}\overline{\nu}\nu$ have to moderately cancel each other. Although the cancellation also requires near degenerate masses and closed $b_L-s_L$ and $b_R-s_R$ mixing angles, the minimal GUT prediction $y_2=y_3$ makes it much less ad-hoc.

\subsection{Emerging flavor mixing angles}
The implementation of flavor mixing still needs to be discussed. 
Although the $b\rightarrow c, s$ transitions generally require sizable non-diagonal Yukawa couplings, Eq.~(\ref{126H}) contains only one Yukawa matrix $Y_{126}$, and one can always choose a flavor diagonal basis. Consequently, no flavor mixing can arise at any scale in the minimal set-up. Thus where does the necessary flavor mixing come from? One may construct extended models that introduce sizable off-diagonal LQ Yukawa couplings while leaving the other sectors unaffected or only mildly modified. Such an approach, nevertheless, is ad hoc and sacrifices the elegance of minimality.

From a different perspective, it is also necessary to include flavor mixing for a complete analysis.  
One physical reason is that the flavor-conserving limit is not protected by gauge symmetry, but comes out as a result of minimality, \textit{i.e.} because the theory contains only one unique Yukawa coupling matrix. Since $M_{\text{GUT}}$ is close to the Planck scale, new particles related to gravity are believed to lie not far beyond $M_{\text{GUT}}$ with additional Yukawa couplings.
Potential flavor violating corrections can arise from additional scalars that couple to the fermions via loops~\cite{Witten:1979nr,Bajc:2004hr, Bajc:2005aq}, or higher-dimensional operators~\cite{Ellis:1979fg} generated by vector-like fermions or other gravity related GUT multiplets. 
A second reason is our new finding in this work that although the flavor-conserving limit is stable under the SM RGEs~\cite{Ma:1979cw, Sasaki:1986jv, Grzadkowski:1987tf}, it becomes unstable when light LQs are introduced. If flavor conservation is not exact at $M_{\text{GUT}}$, the RG flow can amplify deviations and generate larger flavor violating interactions at low energies.\footnote{A similar example is the  amplification of the neutrino mixing angles in the MSSM~\cite{Babu:1993qv, Tanimoto:1995bf, Balaji:2000gd, Balaji:2000ma, Hagedorn:2004ba} by the RGEs. } The flavor mixing can manifest itself as an emergent phenomenon with a small seed at deep UV, while complexity arises dynamically through the evolution towards the IR.

Due to the two reasons mentioned above, we must include the most general and relevant flavor violating effects in a model independent way. The minimal GUT predicts the flavor conserving interaction Eq.~(\ref{yukawaDef}) only --- an over-restrictive result. Despite this, no matter which theory lies behind it, \emph{all} possible 
2nd-to-3rd generation quark flavor violating effects can be captured by the following four new interacting terms: 
\begin{equation}
\label{nondiag}
    \mathcal{L}_Y^{\epsilon}~=~ \epsilon^{ct} y_t \overline{Q_L^2}t_RH_u+ \epsilon^{sb} y_b \overline{Q_L^2}b_RH_d+\epsilon^{bs} y_b \overline{Q_L^3}s_RH_d+ \epsilon^{bs}_1y_1 \overline{s_R^c}\tau_R \widetilde{S}_1+\text{h.c.}
\end{equation}
Here, $Q_L^2=(c_L,s_L)$ denotes the second-generation left-handed quark doublet. The flavor basis is chosen such that the $S_3$ and $\widetilde{R}_2$ couplings are aligned to $Q_L^3$ and $b_R$, respectively, which can be done without loss of generality. 
Once $H_u, H_d$ acquire VEVs the diagonalisation of the quark mass matrix provides the needed flavor-changing couplings for $S_3$ and $\widetilde{R}_2$. 
Since the $t_R-c_R$ rotation remains unphysical, we do not need to include the $\overline{Q_L^3}c_RH_u$ term. 
Mixings involving leptons or first-generation fermions are omitted, as they are irrelevant for  the $b\to s$ and $b\to c$ processes under consideration. 
In the quark sector with only 2nd and 3rd generations, Eq.~(\ref{nondiag}) serves as the maximally possible flavor extension, as long as no other light scalar particles are included. (Our approach to add the maximally allowed flavor-breaking term $\mathcal{L}_Y^{\epsilon}$ is a well-established methodology, 
in analogy with  the soft SUSY-breaking Lagrangian for the MSSM which is also chosen maximal to cover all possible SUSY-breaking scenarios.) Therefore, we do not specify the underlying mechanism of flavor breaking, but instead explore Eq.~(\ref{nondiag}) since it is generic and covers all possible models.

We believe it is important to remark here, that adding general flavor mixing terms in Eq.~(3.9) does not render the GUT scale relation $m_{\tau}=3m_b$ arbitrary. 
This relation is predicted by the $\text{diag}(1,1,1,-3)$ vacuum structure of the Higgs fields in the adjoint representation of $SU(4)_c$, which is the relevant symmetry to unify quarks and leptons. 
$m_{\tau}=3m_b$ originates from the Higgs field with $SU(4)_c\times SU(2)_L\times SU(2)_R$ quantum numbers $(15, 2, 2)$. To change $m_{\tau}=3m_b$  at the  GUT scale, one needs a $(1,2,2)$ Higgs field  or an effective operator containing $(15,1,1)\times(15,2,2)$. 
By contrast, any extensions to the minimal Yukawa sector leads to flavor violation. 
For instance, two $(15,2,2)$ Higgs fields (which can originate from two $126_H$) incorporate flavor but still predict $m_{\tau}/m_b=3$. 
Furthermore, we will argue that the RG evolution which is responsible for the enhancement of flavor mixing does not change $m_{\tau}/m_b$ because of $SU(4)_c$ symmetry.
Contrary to the case of flavor mixing, 
large corrections to $m_{\tau}/m_b$ can only originate from the mass splitting of particles inside the $SU(4)_c$ multiplets, which is realized with TeV-scale leptoquarks in our work, or a non-minimal SO(10)  Yukawa sector  with large couplings, while small perturbations at the GUT scale cannot change this.

\begin{figure}[t!]
    \centering
    \includegraphics[width=0.95\linewidth]{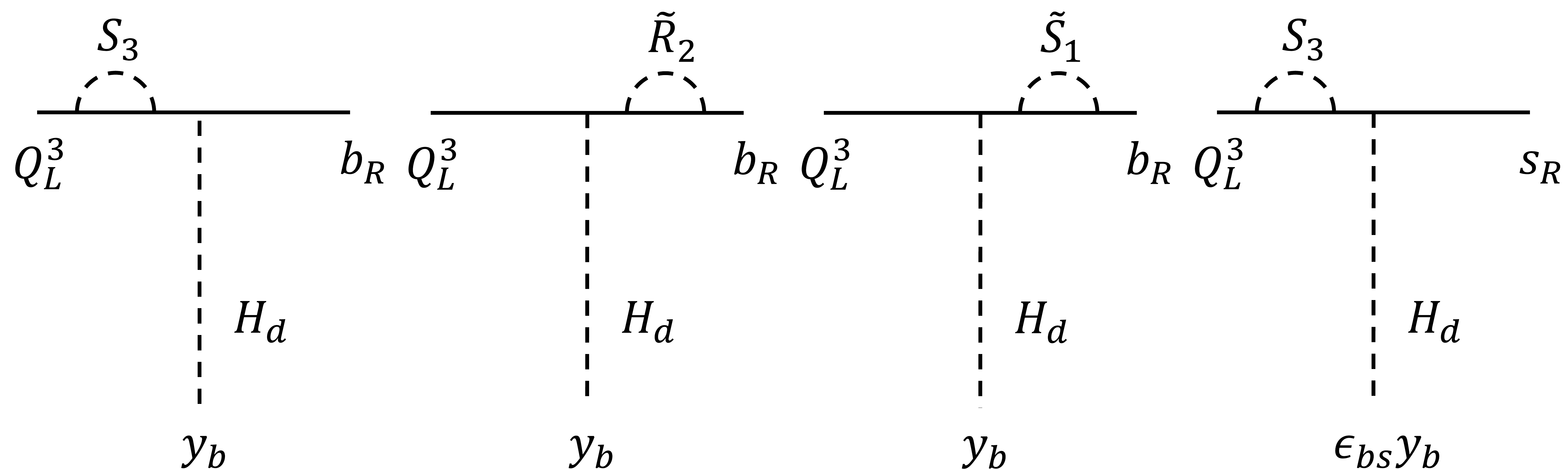}
    \caption{Feynman diagrams illustrating the LQ contributions to the running of $y_b$ and $(\epsilon^{bs}y_b)$. The self-energy correction to $Q_L^3$, arising from $S_3$, is universal  to both $y_b$ and $(\epsilon^{bs}y_b)$. $\widetilde{S}_1$ and $\widetilde{R}_2$ couple to $b_R$ only so only contribute to $y_b$. $s_R$ receives no self-energy corrections.}
    \label{feyndiag}
\end{figure}

Since Eq.~(\ref{nondiag}) contains the general flavor mixing effects, renormalization does not lead to new physical counterterms, ensuring that Eq.~(\ref{yukawaDef}) and Eq.~(\ref{nondiag}) are well defined at all scales.
We then study how the flavor violating corrections evolve from $\Lambda$, the scale at which the new dynamics correcting the minimal GUT ansatz arises (typically $\Lambda\gtrsim M_{\text{GUT}}$),  toward the TeV scale where the light LQs lie.  Taking $\epsilon^{bs}$ as an example, the RG equations from $M_{\text{GUT}}$ to TeV scale are given by:
\begin{equation}
\label{runningeqnondiag}
    \begin{aligned}
    16\pi^2\frac{d}{d \ln{\mu}}(\epsilon^{bs}y_b)~=&~\epsilon^{bs}y_b\left( -\frac{5 g_1^2}{12}-\frac{9 g_2^2}{4}-8 g_3^2+\frac{y_t^2}{2}+\frac{9 y_b^2}{2}+y_{\tau}^2+\frac{3 y_3^2}{2}\right). \\
    \end{aligned}
\end{equation}
Here, we set $\epsilon_1^{bs}=0$ for simplicity, and neglect higher order 
flavor-violating terms. Comparing with the $y_b$ running in Eq.~(\ref{runningEq}), the $\left(\frac{y_1^2}{2}+y_2^2\right)$ term is absent in Eq.~(\ref{runningeqnondiag}). This is because the LQ couplings are third-generation specific and do not contribute to the self-energy corrections for $s_R$, as shown in Figure~\ref{feyndiag}. During the evolution from UV to IR, $(\epsilon^{bs}y_b)$ decreases more slowly than $y_b$, and the mixing parameter $\epsilon^{bs}$ effectively increases. This behavior is further enhanced by the negative terms arising in its explicit RG equation:
\begin{equation}
\label{runningeqmixing}
    16\pi^2\frac{d}{d \ln{\mu}}\epsilon^{bs}=-\epsilon^{bs}\left( \frac{y_1^2}{2}+y_2^2\right). \\
\end{equation}
If $\epsilon^{bs}$ is zero at a given scale, it remains exactly zero for all 
scales $\mu$ since the right-hand side of Eq.~(\ref{runningeqmixing}) vanishes. However, any slight deviation of $\epsilon^{bs}$ from zero would be driven towards sizable values as $\mu$ decreases. As a consequence, the $s_R-b_R$ mixing is enhanced in the IR relative to the UV value by a factor of $\left(\frac{M_{\text{GUT}}}{\text{TeV}}\right)^{\frac{y_1^2+2 y_2^2}{32\pi^2}}$. This leads to a flavor misalignment between the Higgs-Yukawa and LQ sectors. Upon choosing the physical basis in which the $H_d$ Yukawa coupling is diagonal, the enhanced flavor mixings are then transferred to the $\widetilde{S}_1$ and $\widetilde{R}_2$ couplings.\footnote{There are additional contributions to the flavor misalignment. The flavor-violating LQ interaction itself, chosen zero in Eq.~(\ref{nondiag}), also receives radiative correction. The beta-function contains an inhomogeneous term proportional to $\epsilon^{bs}y_b^2$. It is much smaller than the $\epsilon^{bs}\left( \frac{y_1^2}{2}+y_2^2\right)$ contribution, because Eq.~(\ref{initial}) tells that $y_b^2$ is $12$ times smaller than $y_1^2$ or $y_2^2$ at $M_{\text{GUT}}$. For simplicity, we neglect this sub-leading effect.} 
The physical bottom quark mass also receives a correction which is quadratic in $\epsilon^{bs}$ from the diagonalization of the down-type quark mass matrix. Similar corrections also apply to $m_\tau$. Both together may further refine the  $b-\tau$ unification, which is welcome to fix e.g.\ the small mismatches in Fig.~\ref{S3R2} and the $S_3, \widetilde{S}_1$ scenario of Fig.~\ref{modelindepence}.

\begin{figure}
    \centering
    \includegraphics[width=0.65\linewidth]{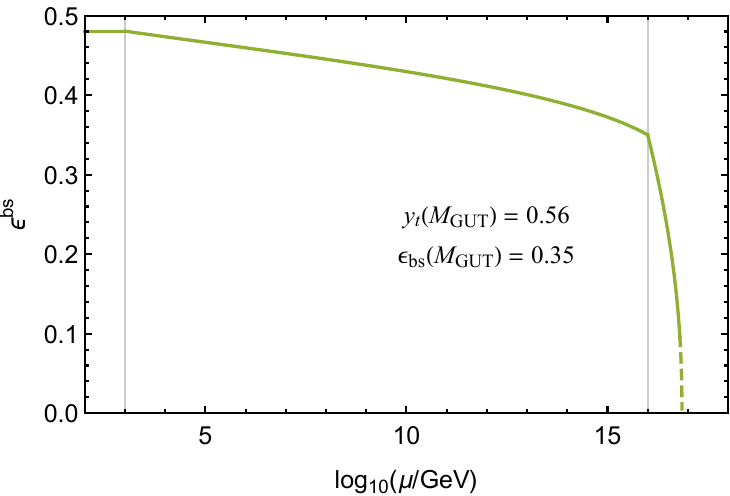}
    \caption{RG evolution of $\epsilon^{bs}$ with $y_t(M_{\text{GUT}})=0.58$. The dashed line lies in the region where the Yukawa couplings become non-perturbative and thus may not reflect the physical reality.}
    \label{mixbs}
\end{figure}

We illustrate our idea in Figure~\ref{mixbs}. Keeping $y_t(M_{\text{GUT}})=0.58$ as the initial condition, the solid line shows the evolution of $\epsilon^{bs}$. The preservation of flavors is clearly not a stable solution of the RG equations, because $\epsilon^{bs}$ increases as $\mu$ decreases. The drawback is that the evolution below $M_{\text{GUT}}$ is too slow: Achieving $\epsilon^{bs}\approx 0.5$ at the TeV scale typically requires its GUT scale value be at least $0.35$, which one may consider to be too large for a perturbation. 
However, we find $\epsilon^{bs}$ changes sufficiently fast above $M_{\text{GUT}}$. One reason is that the $(6_c,1_L,1_R)$, $(\overline{10_c}, 1_L,3_R),$ and $(15_c,2_L,2_R)$ multiplets contained in $126_H$ all contribute to the $b_R$ self-energy above the threshold. We include their effects according to the matching condition and RG equations from Ref.~\cite{Aulakh:2002zr, Meloni:2016rnt, Aydemir:2018cbb}. Another reason is that the absolute value of the Yukawa coupling becomes large. Using the one-loop running equations for gauge and Yukawa couplings above $M_{\text{GUT}}$~\cite{Vaughn:1978st, Vaughn:1981qi, Fang:2025hlc}, we find $Y_{126}$ increases rapidly and approaches the non-perturbative regime when $\mu\gtrsim 10^{17}$ GeV.\footnote{Large $Y_{126}$ induces a sizable $g_{10}$ through two-loop effects, while a sizable $g_{10}$ in turn slows down the running of $Y_{126}$ or can even drive it to decrease. If the scalar self-couplings are excluded, asymptotic freedom can be restored in the deep UV~\cite{Vaughn:1981qi}.} When $\epsilon^{bs}$ lies in the non-perturbative region, we show it with a dashed line and emphasize that this is only for illustration and may not reflect the physical reality. 
It is worth noticing that this enhancement originates from the self-energy corrections to a single $16_F$, which are therefore universal to all 3rd generation quarks and leptons. The RG evolution from about $10M_{\text{GUT}}$ to $M_{\text{GUT}}$ cannot amplify any perturbation to $m_{\tau}/m_b=3$, due to the protection by $SU(4)_c$ symmetry. 
In conclusion, if physics at the scale $\Lambda \gtrsim 10\, M_{\text{GUT}}$ gives a tiny correction to our minimal theory, the RG effects yield sizable flavor mixings in the IR in general, while TeV-scale LQ are still needed to achieve realistic $b-\tau$ mass relationship.




Instead of invoking unknown dynamics above the GUT scale, a more convincing approach is to add more terms to Eq.~(\ref{runningeqmixing}) and change the evolution below $M_{\text{GUT}}$. This can be achieved by requiring more light scalar components in $126_H$, such as di-quarks~\cite{Arnold:2009ay, Giudice:2011ak, Patel:2022nek, Crivellin:2023saq}, which also speed up the running of all Yukawa couplings. As a result, $\epsilon^{bs}$ increases more rapidly
on the path from UV to IR and permits a much lower value at the GUT scale. From a phenomenological perspective, the di-quarks have also been discussed to explain the $b\rightarrow s\ell\ell$ anomalies~\cite{Crivellin:2023saq} and they may furthermore contribute to the interpretation of the observed CP asymmetry in $D_0\rightarrow\pi^+\pi^-$ and $D_0\rightarrow K^+K^-$ decays~\cite{LHCb:2019hro,LHCb:2022lry, Altmannshofer:2012ur, Iguro:2024uuw}. We leave a detailed analysis in this direction for future work.

The behavior of left-handed quark mixing $\epsilon^{sb}$ and $\epsilon^{ct}$ is similar. 
However, there is an additional element ensuring the preservation of the small $V_{cb}\sim -V_{ts}\sim 0.04$ in the RG evolution: $b_L-s_L$ mixing should be well aligned with $t_L-c_L$ mixing. This further requires $\epsilon^{sb}=\epsilon^{ct}$ to a good precision at $M_{\text{GUT}}$, which can be achieved and protected by $SU(2)_R$ symmetry that connects $t_RH_u$ to $b_RH_d$. Since $SU(2)_L$ gauge symmetry stays unbroken until low energies, the evolution for $\epsilon^{sb}$ and $\epsilon^{ct}$ is identical. We think that this feature is worth a general comment:  Explaining $B$ anomalies requires misaligned $b-s$ mixing for $H_d$ and LQ Yukawa couplings. In the SM, however, neither $b_L-s_L$ nor $b_R-s_R$ mixing is physical; the only observable is the difference between $b_L-s_L$ and $t_L-c_L$ mixing. In other words, the SM quark mixing pattern and beyond-SM flavor structure could have intrinsically different origins, with different relevant symmetries.



\section{Conclusion and outlook}
\label{conclu}

In this article, we demonstrate that  TeV-scale LQs addressing the $B$ anomalies can also resolve the wrong $b-\tau$ mass relation in GUTs with simple Yukawa sectors. Among the six types of LQs contained in $126_H$, $\widetilde{S}_1,\widetilde{R}_2, S_3$ can explain the $b\rightarrow c,s$ anomalies while remaining compatible with grand unification. Fixing exact Yukawa unification at $10^{16}$ GeV and $m_{\widetilde{S}_1}=m_{\widetilde{R}_2}=m_{S_3}=1~\text{TeV}$, the running masses $m_t, m_b, m_{\tau}$ can match well with low-scale observations. This is a predictive result because the theory contains only two arbitrary inputs at leading-log level: $\tan\beta$ and $y_{t}(M_{\text{GUT}})$. We also show that scenarios without $\widetilde{S}_1$ or $\widetilde{R}_2$ can still lead to successful third-generation Yukawa unification. This implies that the successful prediction of mass ratios is not tied to a specific choice of LQ types. Although our minimal SO(10) theory predicts no flavor mixing, we demonstrate that the flavor-conserving  solution becomes unstable when perturbed by small flavor-violating terms at the GUT scale (or slightly above) once TeV-scale LQs are introduced.

Our central goal is to show how Yukawa unification is improved by the LQs. Further identifying an explicit example that helps to understand the $B$ anomalies certainly strengthens our approach.
A relevant set of new parameters describe the deviations 
from a minimal Yukawa sector is needed to obtain flavor mixing.  As an important generic result, we find that flavour violation in the LQ couplings increases by RG effects while small flavour mixing in the SM Higgs Yukawa sector stays small under the RG flow by SU(2) symmetry.
Nevertheless, flavor mixing angles large enough to explain the anomalies can emerge from tiny, SM-like flavor violation at the GUT scale only when the $126_H$ spectrum is further modified. In particular, the light colored states --- such as di-quarks and color-octets --- are typically required to further enhance the RG evolution of the 
flavor mixing angles determining the flavor-changing LQ couplings in the IR.
These light colored states living in $126_H$ can simultaneously improve 
successful gauge coupling unification~\cite{Goto:2023qch}, because the light $S_3$ LQ accelerates the running of $g_2$, which must be balanced by a corresponding speed-up of $g_3$ induced by other light colored states.

The successful $b-\tau$ unification and enhancement for flavor mixings raises the question whether TeV-scale LQ can improve the unification of all three generation SM fermions. 
In principle, light LQs can reduce the number of Higgs multiplets needed compared to the usual GUT set-up. 
The TeV-scale theory then contains less free parameters compared to the generic frameworks, and some LQ couplings to the 2nd generation fermions may be related  to known fermion masses and mixings.
This could result in correlations among flavor observables related to lighter fermions, for instance, $g-2$ of the muon and $CP$ violation for $D$ mesons. 
However, second-generation fermion masses  and related observables are not robustly predicted, because they can be significantly changed by tiny corrections to the minimal theory.

Finally we remark that the physical implication of this work goes beyond unified theories. The original Froggatt-Nielsen paper~\cite{Froggatt:1978nt} introduced two distinct ideas  to address the SM fermion mass pattern: (a) RG running effects and (b) UV model building. The idea (a) is further developed using the framework of fixed points~\cite{Pendleton:1980as, Hill:1980sq}, suggesting that the IR, rather than the UV,  
structure of the theory determines the SM flavor parameters. However, the RG equations of the SM alone do not lead to a simple UV flavor structure, which underpins idea (b), UV model building. Thanks to the recent progress at the high-intensity frontier by LHCb and $B$ factories and in lattice QCD calculations, particularly the successful measurements and predictions of $R(D^{(*)})$, we now have data-driven motivations to extend the SM to a larger IR theory with LQs with TeV-scale masses. The situation has changed compared to 20 years ago and the RG approach to the flavor puzzle deserves new attention. Although light LQs bring a  new hierarchy problem, we argue that the RG approach is somehow a more promising path than UV model-building, because the emergence of complexity from simple structures at small distance scales, as we observe in our RG evolution, is a universal phenomenon and not limited to particle physics.
Well-known examples can be found in condensed matter~\cite{Wilson:1974mb} or other complex systems~\cite{Wolfram:1985qr, Wolfram:2020jjc}.


\section*{Acknowledgments}
We are grateful to Howard Haber for useful comments. This research was supported by the Deutsche Forschungsgemeinschaft (DFG, German Research Foundation) under grant 396021762 - TRR 257. X.G. also acknowledges the support by the Doctoral School ``Karlsruhe School of Elementary and Astroparticle Physics: Science and Technology.''


\newpage

\appendix

\section{Model independence}
\label{append}

\begin{figure}[t!]
    \centering
    \includegraphics[width=0.49\linewidth]{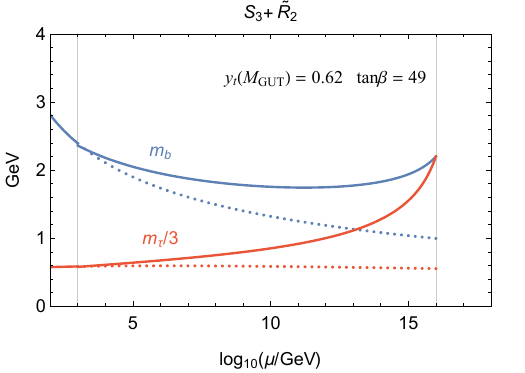}
    \includegraphics[width=0.50\linewidth]{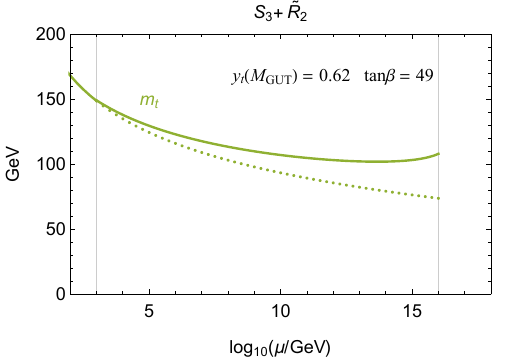}\\
    \vspace{10pt}
    \includegraphics[width=0.49\linewidth]{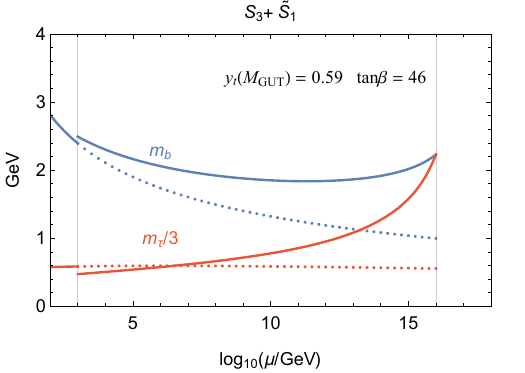}
    \includegraphics[width=0.50\linewidth]{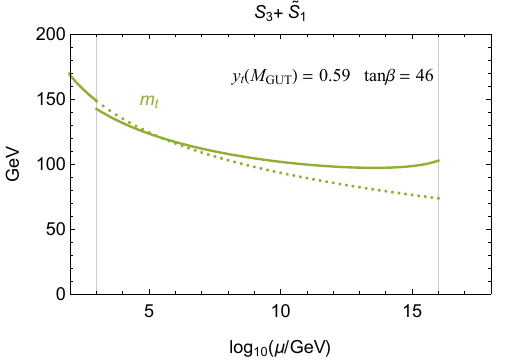}
    \caption{RG evolution in two alternative scenarios to demonstrate model independence. Top: $S_3+\widetilde{R}_2$ with $y_t(M_{\text{GUT}})=0.62,~\tan\beta=49$. Bottom: $S_3+\widetilde{S}_1$ with $y_t(M_{\text{GUT}})=0.59,~\tan\beta=46$. All others setting are same as in Figure~\ref{S3R2}.}
    \label{modelindepence}
\end{figure}

We show scenarios with $S_3+\widetilde{R}_2$ and $S_3+\widetilde{S}_1$ in Figure~\ref{modelindepence} to demonstrate that our results do not depend on the specific choice of LQ type but are rather universal. We take the same technical setup as the three-LQ scenario, but now find successful Yukawa unification for  $y_t(\mu_\text{GUT})=0.62, ~\tan\beta=49, ~y_1\equiv0$ in the $S_3+\widetilde{R}_2$ case, and for $y_t(\mu_\text{GUT})=0.59, ~\tan\beta=46, ~y_2\equiv0$ in the $S_3+\widetilde{S}_1$ case. Interestingly, the $S_3+\widetilde{R}_2$ scenario, which may serve as the minimal LQ model to address the B anomaly, yields a better TeV-scale match than the three-LQ result. In contrast, $S_3+\widetilde{S}_1$ scenario might be unable to fully account for the $R(D^{(*)})$ anomaly without violating the $B\rightarrow K^{*}\overline{\nu}\nu$ constraint, its matching result is also slightly worse. Despite this, all scenarios provide consistent indication that TeV-scale LQs make the $b-\tau$ mass relationship no longer a major concern for minimal SO(10).

\newpage

\bibliographystyle{JHEP}
\bibliography{BanomalySO10.bib}

\end{document}